\documentclass[12pt,prd,floats,showpacs,tightenlines]{revtex4}

\usepackage{amsmath}
\usepackage{amsfonts}
\usepackage{amscd}
\usepackage{epsfig}
\usepackage{amssymb}
\usepackage{tabularx}
\usepackage{graphicx}    

\begin{document}

\title {A gravitational memory effect in ``boosted'' black hole perturbation
theory.} 
\author{Reinaldo J. Gleiser and Alfredo E. Dom\'\i nguez }
\affiliation{ Facultad de Matem\'atica, Astronom\'{\i}a y F\'{\i}sica,
Universidad Nacional de C\'ordoba,\\ Ciudad Universitaria, 5000 C\'ordoba,
Argentina.}
\date{\today}
\email{gleiser@fis.uncor.edu}

\begin{abstract}
Black hole perturbation theory, or more generally, perturbation theory on a
Schwarzschild bockground, has been applied in several contexts, but usually
under the simplifying assumption that the ADM momentum vanishes, namely, that
the evolution is carried out and observed in the ``center of momentum frame''.
In this paper we consider some consequences of the inclusion of a non vanishing
ADM momentum in the initial data. We first provide a justification for the
validity of the transformation of the initial data to the ``center of momentum
frame'', and then analyze the effect of this transformation on the gravitational
wave amplitude. The most significant result is the possibility of a type of
gravitational memory effect that appears to have no simple relation with the
well known Christodoulou  effect.
\end{abstract}


\maketitle

\section{Introduction}\label{sec:intro}

Following its successful application to the close limit of head on collisions,
by Price and Pullin \cite{PrPu}, black hole perturbation theory has become an
important tool in the analysis of the final stages of the coalescence of two
black holes, after their collision or final plunge from their innermost stable
orbit  \cite{PuKy}. In this context, the theory, originally formulated in the
Fourier transformed frequency domain by Regge and Wheeler \cite{ReWhe} and
Zerilli \cite {Zerilli}, is considered in the time domain as a manner of
(perturbatively) solving an initial value problem. Namely, one is given initial
data (a solution of the constraint equations of General Relativity) in the form
of the 3-metric and extrinsic curvature on (some region  of) a three dimensional
spacelike hypersurface $\Sigma$ and the problem is to  find the full metric in
the domain of dependence of the initial data. The type  of data considered here
depends in general on one or more perturbation  parameters $\epsilon_i$, in such
a way that one recovers that for a  Schwarzschild black hole when the
$\epsilon_i$ vanish. The possibility of a perturbative analysis of the evolution
is based on the  expansion of the data in an appropriately chosen angular basis
(tensor  spherical harmonics), and the assumption that the angular component
coefficient  functions can be further expanded in powers of the $\epsilon_i$.
The non trivial part of the Einstein equations can then be cast in the form of
an infinite set of coupled partial differential equations for functions of two
variables (conventionally, the Schwarzschild coordinates $t$ and $r$), that can,
in  principle, be integrated order by order in the perturbation parameters
\cite{physrep}. This, however, does not take into account the invariance of the
geometry under coordinate changes. In fact, when this invariance is fully
accounted for, the relevant physical information ends up being encoded in two
sets of functions, where each element corresponds to an angular mode,
introduced respectively by Regge and Wheeler \cite{ReWhe}, and by Zerilli
\cite{Zerilli}, that satisfy wavelike equations in $t,r$.

While the above analysis, leading to the emergence of the Regge - Wheeler and
the Zerilli functions is based on the general invariance of the theory under
coordinate transformations, in recent applications to the black hole binary
problem one is interested in the evolution of a given initial data, posed on a
given hypersurface $\Sigma$. In perturbation theory, a particular choice is
made of the zeroth order (in the $\epsilon_i$) coordinates. This still leaves
the freedom of choice of coordinates within $\Sigma$, which can be redefined,
provided this introduces changes of the same order as the perturbations
\cite{moncrief}. This {\em gauge} freedom  on $\Sigma$ is very important because
it allows to cast the evolution problem (and corresponding initial data) in
different physically equivalent forms, that simplify either the mathematical
treatment or the physical interpretation \cite{physrep}.

However, in considering more general coordinate transformations one is faced
with the following problem. Suppose that for certain choice of coordinates
$x^{\mu}=(t,r,\theta,\phi)$, $\Sigma$ is the hypersurface $t=0$, and that the
hypersurfaces $\Sigma_t$, corresponding to constant values of $t$ define a
foliation of the space time manifold $\cal{M}$, as would be appropriate for the
evolution of initial data on $\Sigma$, using $t$ as the ``time'' parameter. We
may introduce now a different coordinate system, say $\tilde{x}^{\mu} =
(\tilde{t},\tilde{r},\tilde{\theta},\tilde{\phi})$, such that the constant
$(\tilde{t}$ hypersurfaces provide a different foliation of $\cal{M}$, but where
the transformation between $\tilde{x}^{\mu}$ and $x^{\mu}$  depends on the
$\epsilon_i$ in such a way that we get $\tilde{x}^{\mu}  ={x}^{\mu}$ for
$\epsilon_i=0$. It is not immediate that this more general case  can be treated
as that where we consider coordinate changes only within  $\Sigma$. This is
because, by definition, the initial value problem, and  corresponding initial
data, for the coordinates $\tilde{x}^{\mu}$ correspond  now to constant
$\tilde{t}$, but the restrictions on the limits of the  $\epsilon_i=0$, do not
necessarily imply that, for finite $\epsilon_i$, even if  the hypersurfaces
$\Sigma$ and $\tilde{\Sigma}$, corresponding respectively to  $t=0$ and
$\tilde{t}=0$ intersect at some points, they cannot become widely  separated in
time as we move away from these points. (See below for more  details). But this
would imply, in principle, that in this case the only way we  can implement the
coordinate transformation is if we have already solved the  evolution equations,
since a finite amount of time is required to move in  general from a point in
$\Sigma$ to a point in $\tilde{\Sigma}$.

The previous discussion is relevant to the following problem. Suppose we  are
given a family of initial data for the Einstein equations, (3-metrics $g_{ij}$,
extrinsic curvatures $K_{ij}$), depending smoothly on a parameter $P$, such
that for $P=0$ we recover Schwarzschild's space time, and for $P\neq 0$ the
initial data is asymptotically flat, with ADM momentum $P$. Quite generally,
for small $P$, we expect this data to evolve into a ``boosted'' Schwarzschild
black hole, possibly accompanied with the emission of a certain amount of
gravitational radiation. However, if we consider applying the
Regge-Wheeler-Zerilli perturbation theory to analyze this evolution, we are
faced with the problem that this theory is based on the assumption that the
evolution leads to a {\em stationary} black hole, essentially centered at the
origin of spatial coordinates, while the ``natural'' evolution of the given
initial data leads to a {\em non stationary} final state of the black hole.
Clearly, this problem could be solved by choosing a new foliation, where the
final state of the black hole is stationary, but such transformation, for any $P
\neq 0$ involves a ``Lorentz boost'', with arbitrarily large separations of the
hypersurfaces $\Sigma$ and $\tilde{\Sigma}$, as can be seen by considering the
simpler similar case in Minkowski spacetime.

The evolution of conformally flat ``boosted'' single black hole initial data in
black hole perturbation theory, was analyzed in a recent paper by Khanna,
Gleiser and Pullin \cite{KhGlPu}, (referred as I in what  follows), but there
the previous problem was given only a heuristic treatment.  In more detail, the
analysis performed in I makes use of second order  perturbation theory, not as
regards evolution, but rather to carry out a second  order gauge transformation
that eliminates the first order terms, leaving only  the second order
contributions, which then satisfy linearized Einstein  equations. Although the
results obtained regarding radiate wave forms are  qualitatively in agreement
with other results obtained in perturbation theory,  i.e., they show, for
instance, the expected ``quasi-normal ringing'', an  intriguing feature that
distinguishes this case is that the Zerilli function  $\psi$ does not vanish for
large radial distance, approaching instead a  constant non vanishing value.
Since the  radiated energy depends on the time  derivative of $\psi$, the
presence of this constant does not in itself mean  that there is a divergence,
but this behavior is in clear contrast with that  previously observed in other
applications of perturbation theory, and,  therefore, it justifies a more
detailed analysis and interpretation.

The other point that also needs consideration in detail, for the reasons
mentioned above, is the type of gauge transformation performed in I on the
initial data. Its effect was equivalent to a coordinate transformation where
one moves from a slice where the black hole has non vanishing linear momentum,
to a frame where it is ``at rest'', that is, the transformation is essentially
a ``boost''. But this introduces, at least in principle, a transformation that
requires the knowledge of the evolution of the initial data from the
``boosted'' frame to the ``rest'' frame, and therefore, as indicated, it is not
simply equivalent to a relabeling of points on the initial data surface.

In this paper we consider again the problem from a more general point of view.
We first present a justification for the validity of the ``passage to the
center of mass system'' used in \cite{KhGlPu}, and then show that the somewhat
unexpected asymptotic behavior of the Zerilli function found there can be
interpreted as a gravitational memory effect, that might be present in some
form in any problem where one has ``single boosted black hole'' type of initial
data.

\section{A digression on coordinate and gauge transformation}

The development of higher order perturbation theory given in \cite{physrep},
(see also \cite{Bruni}) is  based on the existence of a family of solutions of
Einstein's equations, depending on the parameter $\epsilon$, which includes the
Schwarzschild metric for $\epsilon=0$, and on the possibility of performing
general coordinate transformations, which may also be classified in one
parameter families, with the same parameter, $\epsilon$, as the family of
metrics. It is then assumed that both the metric coefficients, and coordinate
transformation functions may be expanded in powers of $\epsilon$, around
$\epsilon =0$, which naturally leads to a classification in ``orders'', in
accordance with the corresponding power of $\epsilon$ in the expansions. With
these assumptions one obtains an infinite set of relations between the metric
coefficients corresponding to the same geometrical metric, expanded in
powers of $\epsilon$, but written in different coordinate systems, and
the expansion coefficients of the coordinate transformation functions,
each member of the set corresponding to a given order in $\epsilon$. We
 generally call ``nth order gauge transformation'' the relations obtained
 equating coefficients of nth order in $\epsilon$.

For the purpose of applications of perturbation theory, it is only practical to
consider the lowest orders. In particular, we may consider coordinate
transformations that contain only linear terms in $\epsilon$. These naturally,
generate first order gauge transformations, which are linear in the coordinate
(``gauge'') transformation functions, but they also generate higher order gauge
transformations, through terms that are quadratic, cubic, etc, in the gauge
functions. Similarly, we may consider coordinate transformations that are
quadratic in $\epsilon$, to start with. These generate second, fourth, etc,
order gauge transformations, but do not affect the first order terms. Thus we
may consider, as in \cite{physrep}, a sequence of gauge transformations, where
the order of the gauge transformation is raised as we move along the sequence.

In all these considerations, we are assuming that the metric is known in some
four dimensional region of the spacetime manifold, so that the coordinate
transformations are quite general. We remark, however, that an important set of
applications to black hole physics is based on the perturbative solution of an
initial value problem. This requires the introduction of some foliation of
spacetime that singles out a one parameter family of spacelike hypersurfaces,
on one of which the initial data is given. The simplest way of specifying the
hypersurfaces is by introducing a ``time'' coordinate $t$, such that $t=
\mbox{constant}$, on each hypersurface, and the initial data is given for
$t=0$. Under a general gauge transformation of the kind described above, we
might introduce a new ``time'' coordinate $t'$, such that it also provides a
foliation of spacetime, and the initial value problem could, in principle, be
solved starting with initial values on the hypersurface $t'=0$.

We notice, however, that the initial value problem, from its formulation,
implies in practical applications that the metric (and its first $t$
derivative) is known only for $t=0$, on some ``given'' hypersurface. If this is
all the knowledge of the metric that we have at the beginning, and, if the
hypersurfaces $t=0$, and $t'=0$, do not coincide, to obtain the corresponding
initial value for $t'=0$, we need in principle to solve the evolution
equations, since points on $t'=0$, may be arbitrarily far to the future, or
past, of points on $t=0$. Thus, in practice, where an initial value problem is
concerned, one cannot apply the full set of gauge transformations, but a
restriction must be imposed  so that either, there is no change in the initial
data hypersurface, or the change is only of the order of $\epsilon$ considered.

\subsection{A toy model}

We may illustrate these points with a ``toy model''. Consider a field theory in
one-plus-one dimensions, where the field satisfies the Sine-Gordon equation,
\begin{equation}
\label{S-G1}
{\partial^2 \phi(x,t) \over \partial t^2} - {\partial^2 \phi(x,t)
\over \partial x^2} +\sin \phi(x,t) = 0
\end{equation}
Any solution $\phi(x,t)$ of (\ref{S-G1}) is completely determined by the
``initial data'', $$\phi(x,t=0)\;\;,\;\; (\partial \phi(x,t) / \partial
t)|_{t=0} \;.$$  Equation (\ref{S-G1}) is invariant under transformations
(``boosts'') of the form,
\begin{eqnarray}
\label{boo01}
x & = & \cosh(\eta) x' - \sinh(\eta) t' \nonumber \\
t & = & \cosh(\eta) t' - \sinh(\eta) x'
\end{eqnarray}
if $\phi(x,t)$ transforms as a scalar, i.e., $\phi'(x',t') =
\phi(x(x',t'),t(x',t'))$. Moreover, corresponding to any solution $\phi(x,t)$ of
(\ref{S-G1}), we may define the quantities,
\begin{eqnarray}
\label{boo02}
E & = &  \int{[(\phi,_t)^2+(\phi,_x)^2+\sin^2 \phi] dx} \nonumber \\
P & = & \int{(\phi,_t) (\phi,_x) dx}
\end{eqnarray}
where the integrals are computed for constant $t$, but $E$, and $P$, are
actually independent of $t$, and transform as the $(t,x)$ components of a
two-vector, i.e., under the transformation (\ref{boo01}), we have,
\begin{eqnarray} \label{boo03}
P  & = & \cosh(\eta) P' - \sinh(\eta) E'
\nonumber \\ E  & = & \cosh(\eta) E' - \sinh(\eta) P'
\end{eqnarray}
where $E',P'$ are related to $\phi'$ as $E,P$ are related to $\phi$. We may
think of $E$ and $P$, as the ``observable'' quantities to be computed from
$\phi$. The crucial point here is that $E$ and $P$ may be computed solely in
terms of the initial data for $t=0$, and similarly $E'$ and $P'$ may be
computed in terms of the initial data for $t'=0$.

To make contact with the discussion in this paper, we may view the
transformations (\ref{boo01}) as defining a one parameter set of solutions of
 (\ref{S-G1}), namely, we define
\begin{equation}
\label{boo04}
\phi(x,t,\eta) = \phi_o(\cosh(\eta) x - \sinh(\eta) t,
\cosh(\eta) t - \sinh(\eta) x )
\end{equation}
where $\phi_o(x,t)$ is some solution of (\ref{S-G1}). Correspondingly, we have a
one parameter set of 2-vectors $(E(\eta),P(\eta))$, obtained from
$\phi(x,t,\eta)$ through (\ref{boo02}).

Suppose now that we want to consider the evolution equation (\ref{S-G1}) for the
different $\phi(x,t,\eta)$ as an initial value problem.  If the solution
$\phi_o(x,t)$ is given, (essentially for {\em all} $t$), then we have,
\begin{eqnarray} \label{boo05}
\phi(x,0,\eta) & = & \phi_o(\cosh(\eta) x  ,  -
\sinh(\eta) x ) \nonumber \\   \phi(x,0,\eta),_t   & = & -  \phi_o(x,0),_x
\sinh(\eta) +   \phi_o(x,0),_t   \cosh(\eta)
\end{eqnarray}
and, somewhat trivially, we can solve the initial value problem for
the data $(\phi(x,0,\eta), \phi(x,0,\eta),_t )$, to recover $\phi(x,t,\eta)$,
and $(E(\eta),P(\eta))$.

On the other hand, if all that is known for  $\phi_o(x,t)$ is the initial data
for $t=0$, then, we cannot write the right hand side of (\ref{boo04}), and we
need to solve the initial value problem for $\phi_o(x,t)$ before we can
proceed. However, if we are only interested in an ``infinitesimal'' boost,
i.e., the limit $\eta \rightarrow 0$, we may attempt a computation of the right
hand sides of (\ref{boo04}) by expanding $\phi_o$ in a power series in $\eta$.
Namely, since, \begin{eqnarray}
\label{boo01a}
x & = &  (1+\eta^2/2+ ...) x' -  (\eta + ...) t'
 \nonumber \\ t & = &  (1+\eta^2/2+ ...) t' -  (\eta + ...) x'
\end{eqnarray}
we have
\begin{eqnarray}
\label{perturb01}
\phi(x,0,\eta) & = & \phi_o((1+\eta^2/2+ ...) x, -  (\eta + ...) x) \nonumber \\
& \simeq & \phi_o(x,0) -  \phi_o(x,0),_t \eta \; x + ...
\end{eqnarray}
and a similar expression for $\phi(x,0,\eta),_t $. But, due to the presence of
the factors $x$ on the right hand, this means that by restricting to the lowest
powers of  $\eta$, we may obtain an expression for the initial data for
$\phi(x,t,\eta)$  that differs drastically from the exact form. We notice, for
instance, that  while the exact initial data might be square integrable, this
might not hold  for the ``perturbative'' expression in the right hand side of
(\ref{perturb01}). The problem here may be traced to the non commutativity of
the limits $|x'| \rightarrow \infty$, and $\eta \rightarrow 0$. In the ``boost''
interpretation,  for any $\eta \neq 0$, the "hypersurfaces" $t=0$, $t'=0$,
become arbitrarily  separated at large $|x'|$ (or $|x|$), and the evolution
equations must be  satisfied to move from one to the other.

This does not mean that the ``boost'' transformation cannot be used in a
``perturbative'' sense in any Lorentz invariant model. Consider, instead of the
general form for $\phi_o(x,t)$  the ``solitons'',
\begin{equation}
\label{soliton}
\chi(x,t,\beta) = 4 \arctan(\exp(\cosh(\beta)x -\sinh(\beta)t))
\end{equation}
where $\beta$ is a constant.

Let us take $\phi_o(x,t) = \chi(x,t,\beta=0)$. This solution is {\em static},
 i.e., independent of $t$, and the solutions for $\beta \neq 0$ are obtained
 applying a boost with  $\eta =  \beta$ to $\phi_o(x,t)$. In particular, the
 initial data for $\phi(x,t,\eta)$ will be related to $\phi_o(x,t)$ by
\begin{eqnarray}
\phi(x,0,\eta) & = & \phi_o(\cosh(\beta)\;x,0) \nonumber \\
& = & 4 \arctan(\exp(\cosh(\eta)x)) \nonumber \\
\phi(x,0,\eta),_t & = & -
\phi_o(\cosh(\beta)\;x,0),_x \; \sinh(\beta)/\cosh(\beta)\nonumber \\
& = & -2 \sinh(\eta)/\cosh(\cosh(\eta)\;x)
\end{eqnarray}
and we notice that the ''initial data'' for $\phi_o(x,t)$ is enough to
compute that for $\phi(x,t,\eta)$, for all $\eta$.

If we consider again an ``infinitesimal'' boost ($ \eta \rightarrow 0$) of the
static soliton, we find
\begin{equation}
\phi' \simeq 4 \arctan(A\exp(\cosh(x)) - 4 { A t v \over A^2\exp(x)+\exp(-x)} +
 O(v^2)
\end{equation}
and we notice that in this case the ``perturbation'' is uniformly bounded in
$x$, and we may, for instance, use this expression for $\phi'$ to compute, say,
$E$ or $P$, to the corresponding order in $\eta$. This behavior is quite
different from that found for a general non static solution, and originates in
the fact that the ``unperturbed'' solution is static, (independent of $t$). If
we repeat the arguments for the failure of the expansion in the general case,
we notice that for the static solution, the initial data is the same as the
solution for {\em all} $t$, and therefore the initial data on a ``boosted''
slice is simply obtained by ``boosting'', i.e. changing coordinates in the
initial data on the static slice.

Going back to the black hole perturbation problem, we notice that the
Schwarzschild metric, in appropriate coordinates, is manifestly static, and, 
precisely for  the same reason as above, a coordinate transformation equivalent
to a  ``boost'', when applied to this metric, although changing the initial data
slice, is equivalent to a gauge transformation that does not change the slice,
because the evolution equations are automatically satisfied. This will be
considered in detail in the following Sections.

\section{Review of $\ell=1$ even parity linear perturbations}

It will be useful to review some properties of $\ell=1$ even parity linear
perturbations of a Schwarzschild black hole \cite{ReWhe,Zerilli,KhGlPu}.
Restricting to axisymmetry, these may be written in
general in the form,
\begin{eqnarray}
\label{ell1even}
g_{t t}^{(1)} & = & (1-2M/r) H^{(1)}_0(t,r) \cos(\theta) \nonumber \\
g_{r t}^{(1)} & = &  H^{(1)}_1(t,r) \cos(\theta) \nonumber \\
g_{r r}^{(1)} & = & 1/(1-2M/r) H^{(1)}_2(t,r) \cos(\theta) \nonumber \\
g_{r \theta}^{(1)} & = & -h^{(1)}_1(t,r) \sin(\theta) \nonumber \\
g_{\theta t}^{(1)} & = & -h^{(1)}_0(t,r) \sin(\theta) \nonumber \\
g_{\theta \theta}^{(1)} & = & r^2 K^{(1)}(t,r) \cos(\theta) \nonumber \\
g_{\phi \phi}^{(1)} & = & r^2 \sin^2(\theta) K^{(1)}(t,r) \cos(\theta)
\end{eqnarray}

One can show that the general solution of the (linearized) vacuum Einstein
equations satisfied by these perturbations  can be written in the form,
\begin{eqnarray}
\label{ell1even2}
 {H_0^{(1)}}(t,r) & = & - {2 M\over r (r-2M)} {M_1^{(1)}}(t,r) - 2\,{\frac
{\partial }{\partial t}}\,  {M_0^{(1)}}(t,r) \nonumber \\
 {H_1^{(1)}}(t,r)& = & {  r\over (r-2M)}\,{\frac {\partial
}{\partial t}}\,{M_1^{(1)}}(t,r)   -
{ (r - 2\,M)\over r} \,{\frac {\partial }{\partial r
}}\,{M_0^{(1)}}(t,r) \nonumber \\
{H_2^{(1)}}(t,r) & = & - {2 M\over r (r-2M)} {M_1^{(1)}}(
t,r)   + 2\,{\frac {\partial }{\partial r}}\,
{M_1^{(1)}}(t,r)\nonumber \\
{h_0^{(1)}}(t,r)& = & r^{2}\,{\frac {\partial }{\partial t}}\,
{M_2^{(1)}}(t,r) - { (r - 2\,M)\over r}
{M_0^{(1)}}(t,r) \nonumber \\
{h_1^{(1)}}(t,r) & = &  {r \over r-2M} {M_1^{(1)}}(t,r)
  + r^{2}\,({\frac {\partial }{\partial r}}\,
{M_2^{(1)}}(t,r))\nonumber \\
{K^{(1)}}(t,r) & = & {2 \over r} M_1^{(1)}(t,r)
   - 2\,M_2^{(1)}(t,r)
\end{eqnarray}
where $M_0^{(1)}$, $M_1^{(1)}$ and $M_2^{(1)}$ are arbitrary functions of
$(t,r)$.

Conversely, given any solution of the (linearized) vacuum Einstein equations,
the functions $M_i^{(1)}$ are given by,
\begin{eqnarray}
 M_0^{(1)} & = & - { r \over (r-2M)}\left(
 {h_0^{(1)}}  + { {r^2} \over {2}}{K}^{(1)},_t\right)
  +  {r^2\over 6M}
   \left( r^{2} {K}^{(1)},_r - r\, {H_2^{(1)}}
 + 2\, {h_1^{(1)}} \right)_{,t}
 \nonumber \\
 {M_1^{(1)}} & = &{r-2M \over 6 M}
 \left(r^{2}\,
 {K}^{(1)},_r - r\, {H_2^{(1)}}  + 2\, {h_1^{(1)}}
 \right)\nonumber \\
M_2^{(1)}  & = & - { 1 \over 2}
 {K^{(1)}}    + {r-2M \over 6 r M}
 \left(r^{2}\,
 {K}^{(1)},_r - r\, {H_2}^{(1)}  + 2\, {h_1^{(1)}}
  \right)
\end{eqnarray}

An important property of the $\ell=1$ even parity perturbations is their
relation to the ADM linear momentum $p_{\mu}$ associated to an asymptotically
flat initial data set (3-metric $h_{\mu\nu}$ and extrinsic curvature
$K_{\mu\nu}$) given on an asymptotically flat hypersurface $\Sigma$. To obtain
expressions more directly related to black hole perturbation theory, we
consider asymptotically Euclidean (cartesian) coordinates $\{x^i\}$, and a
related set of spherical coordinates $\{r,\theta, \varphi\}$, which we indicate
by $\{x^{\mu}\}$. One can show that the Euclidean components of $p_{\mu}$ are
given by
\begin{equation}
\label{eqP1}
p_i= {1 \over 8 \pi} \lim_{r \rightarrow \infty}
\sum_{\mu,\nu = 1}^3 \int  {\partial x^{\nu} \over \partial x^i} \left(
K_{\mu\nu}n^{\mu} -
K^{\mu} {}_{\mu} n_{\nu}\right)
  \; d A
\end{equation}
where $i,j,...$ refer to Euclidean (cartesian) coordinates and $\mu,\nu...$ to
spherical coordinates. The integral is taken over a sphere of radius $r$, with
$n^{\mu}$ the unit normal to the sphere and $dA = r^2 \sin \theta \, d \theta
\, d \varphi$, the area element on the same sphere.

We recall that if we write the 4-metric in the form
\begin{equation}
ds^2 = -N^2 dt^2 + h_{ij} (dx^i +N^i dt) (dx^j +N^j dt)
\end{equation}
where $x^i$ are coordinates on the hypersurface, and $N$ and $N^i$ are
respectively the {\em lapse} function and {\em shift} vector, we have
\begin{equation}
{\partial h_{ij}\over \partial t} =  N_{i|j} +N_{j|i} - 2 N K_{ij}
\end{equation}
where  ``$|$'' indicates covariant derivative with respect to 3-metric $g_{ij}$,
on the hypersurface $\Sigma$. Then, it is easy to show that if the metric is
given in the form (\ref{ell1even}), and $\zeta$ is a perturbation parameter upon
which the $\ell =1$ even perturbation depend linearly, equation
(\ref{eqP1})takes the form,
\begin{equation}
\label{eqPz}
P_z=\lim_{r \rightarrow \infty}{ \zeta r \over 6} \left[{H_1^{(1)}}+
{\partial {h_1^{(1)}} \over \partial t}
 - r {\partial {K^{(1)}} \over \partial t}
 - {\partial {h_0^{(1)}} \over \partial r}\right]
\end{equation}

We remark, more generally, that only the $\ell=1$, even parity perturbations
contribute asymptotically to $P_z$.

\section{The effect of a boost on the Zerilli function}

Suppose we have a black hole perturbation problem where the leading
perturbations, which we take as of order $\epsilon$, (where $\epsilon$ is some
parameter), are of even parity with $L=2$, $m=0$. Assume further that the metric
is written in an asymptotically flat gauge, and that first order perturbation
theory is appropriate for an analysis of the evolution of the perturbations.
This implies that both the linear and angular momentum vanish to order
$\epsilon$. We may construct the corresponding ($\ell=2$, $m=0$, even parity)
Zerilli-Moncrief function which may be written as,
\begin{equation}
\label{funcZer1}
\psi(t,r) = {r \over 3 } \left[
  {(r-2M)\over (2r+3M)} \left(  {H^{(2)}_2 }  - r {K^{(2)}},_r   +
3\,r {G^{(2)}},_r  -
{6\over r} {h^{(2)}}_1  \right) +
 {K^{(2)}}  \right]
 \end{equation}
and the vacuum Einstein equations imply that $\psi$ satisfies the Zerilli
equation,
\begin{equation}
\label{eqZer1}
{\partial^2 \psi \over \partial t^2}-{\partial^2 \psi \over \partial r_*^2} - V
\psi =0
\end{equation}
where $r_* = r +2M \ln(r-2M)$, and, for even parity $\ell=2$, the ``potential''
$V$ is given by,
\begin{equation}
\label{potential}
V(r) =6 \left(1-{2M \over
r}\right){(4r^3+4Mr^2+6rM^2+3M^3) \over r^3 (2r+3M)^2}
\end{equation}

Suppose further that on some constant $t$ slice we have $\psi \longrightarrow 0$,
and $(\partial \psi /\partial t) \longrightarrow 0$ as $r \longrightarrow
\infty$, with $\psi$ and $(\partial \psi /\partial t)$ some smooth
bounded functions for $r \geq 2M$. Then using (\ref{eqZer1}) we
generally find that, after an eventual ``quasi-normal
ringing'' type of waveform, $\psi \longrightarrow 0$, for all fixed $r>>2M$,
as $ t\longrightarrow \infty$. This in turn implies that, given the simple
relationship between $\psi$ and the gravitational wave amplitude, an
interferometric detector,  initially in a certain state, will return to that
state after the passage of the quasinormal ringing wave, with no ``memory''
effect. Examples of this type can be found, e.g., in \cite{PrPu}. We also notice
that generally, if the initial data contains only even $L$ terms, no odd $L$
values will be generated in the evolution.

Let us assume now that $\epsilon>0$. We may set $\epsilon =P^2$, and consider
$P$ as a perturbation parameter. Then, in the spirit of \cite{ReWhe}, we assume
that a {\em full solution} of Einstein's equations exists, given by a metric
$g_{\mu\nu}(r,\theta,\phi,t)$, for which $\psi$ is obtained by an appropriate
expansion in $P$. Moreover, we may assume that this metric is given in a
coordinate system where it is explicitly in asymptotically flat form, and that
for $P=0$ we recover the Schwarzschild form for the metric. We may now consider
applying on $g_{\mu\nu}$ a {\em coordinate transformation} of the form,
\begin{eqnarray}
\label{transf01}
t & = & t'+P {\cal{M}}_0(t',r') \cos(\theta') \nonumber \\
r & = & r'+P {\cal{M}}_1(t',r') \cos(\theta') \nonumber \\
\theta & = & \theta'-P {\cal{M}}_2(t',r') \sin(\theta') \nonumber \\
\phi & = & \phi'
\end{eqnarray}

Then, (recall the previous discussion) if we expand the transformed metric in
powers of $P$ near $P=0$, we find that the zeroth order terms take again the
static Schwarzschild metric form in $(r',\theta',\phi',t')$ coordinates, while
to order $P$ we find $\ell=1$, even parity terms of the form
(\ref{ell1even}), with the coefficients given in terms the ${\cal{M}}_i$ by
expression of the form (\ref{ell1even2}), where the functions ${{M}}_i$
are replaced by ${\cal{M}}_i$. This is quite general, but, to simplify the
physical interpretation, we may now restrict the functions ${\cal{M}}_i$ in such
a way that the resulting $\ell=1$, even parity terms satisfy the conditions of
asymptotic flatness. By this we mean that asymptotically for large $r'$ we
should have ${H_0^{(1)}}(t',r')$, ${H_2^{(1)}}(t',r')$  and ${K^{(1)}}(t',r')$
at most of order $1/{r'}^2$, and ${H_1^{(1)}}(t',r')$, ${h_0^{(1)}}(t',r')$  and
${h_1^{(1)}}(t',r')$ at most of order $1/{r'}$. One can then check that this is
possible only if the functions ${\cal{M}}_i$ are asymptotically of the form,
\begin{eqnarray}
\label{restric1}
 {\cal{M}}_0^{(1)} & = & a_0 r' +a_1 +a_2/r' + {\cal{O}}(1/{r'}^2)
 \nonumber \\
 {\cal{M}}_1^{(1)} & = & a_0 t'+ a_3 + {\cal{O}}(1/{r'})\nonumber \\
{\cal{M}}_2^{(1)}  & = & (a_0 t'+a_3)/r'+((a_1-2 M a_0) t'+a_4)/{r'}^2
 +{\cal{O}}(1/{r'}^3)
\end{eqnarray}
where the $a_i$ are constants. Moreover, if we replace the resulting expressions
for ${H_1^{(1)}}(t',r')$, ${h_0^{(1)}}(t',r')$,
${h_1^{(1)}}(t',r')$ and ${K^{(1)}}(t',r')$ in (\ref{eqPz}) we find,
\begin{equation}
P_z=\lim_{r \rightarrow \infty} P \left( M a_0 +{\cal{O}}(1/{r'})
\right]
\end{equation}

Therefore, if we choose,
\begin{equation}
\label{restric2}
a_0=1/M ,
\end{equation}
the $\ell=1$ even parity part of the metric describes an ADM momentum equal to
$P$, and the transformation (\ref{transf01}) carries the metric to a ``boosted
frame'', with momentum $P$.

The preceding discussion is nevertheless incomplete. The reason is that actually
equation (\ref{eqPz}) and its physical interpretation hold only on an
asymptotically flat frame, and one can check that the transformation
(\ref{transf01}), even with the restrictions (\ref{restric1}), (\ref{restric2})
introduces terms of order $P^2$ in the $\ell =0$ and $\ell=2$ even parity part
of the metric that are not compatible with (explicit) asymptotic flatness.
This, however, does not modify the interpretation of $P$ as the ADM momentum.
The reason is that we can restore explicit asymptotic flatness introducing a new
coordinate (gauge) transformation, of order $P^2$, that involves only the
$\ell=0$ and $\ell=2$ even parity terms. This has no effect on the
even parity, $\ell=1$ terms and, therefore, leaves the right hand side of
(\ref{eqPz}) {\em unchanged.}

Summarizing, we see that to change from the initial asymptotically flat slicing
with vanishing ADM linear momentum to a new asymptotically flat slicing where
this takes the value $P$, we need to perform at least the equivalent of a gauge
transformation of order $P$, followed by one of order $P^2$ that has no effect
on the ADM momentum.

But now we are in position to discuss the relationship between the
Zerilli-Moncrief function $\psi$ given by (\ref{funcZer1}) in the zero momentum
frame, and the corresponding quantity $\psi_B$ constructed using equation
(\ref{funcZer1}), with the functions on the right hand side as given in the
momentum $P$ frame. These two functions are not equal but, considering now the
effect to order $P^2$ of the transformation (\ref{transf01}), (i.e., as a second
order gauge transformation) one can verify that $\psi_B - \psi $ can be written
as a quadratic homogeneous expression involving {\em only} the ${\cal{M}}_i$ and
their $t'$ and $r'$ derivatives. The explicit expressions are rather lengthy,
but the important result is that with the restrictions (\ref{restric1}),
(\ref{restric2}) we find that for large $r'$,
\begin{equation}
\label{Deltapsi}
\psi_B - \psi = - {2 P^2 \over 3M} +{(17 -16 a_1 -a_1^2) P^2 \over 9 r'} +
{\cal{O}}(1/{r'}^2)
\end{equation}
Therefore, for large $r$ and any finite $t$, $\psi_B - \psi$ approaches
the constant value $-2 P^2 /3M$ irrespective of the details of the $\ell=2$
data. The crucial point that makes this result non trivial is that $\psi_B -
\psi$ is {\em not} changed by the subsequent transformation of order $P^2$ that
restores asymptotic flatness, because the expressions for both $\psi$ and
$\psi_B $ are {\em gauge invariant} under those transformations.

An immediate consequence of these results for initial data sets
$\left\{ {g_B}_{ij},{K_B}_{ij}\right\}$ with non vanishing linear ADM
momentum $P$, is that if the function $\psi_B $ vanishes for large $r$, then
the corresponding Zerilli function in the ``center of momentum'' frame (i.e.
the frame with vanishing ADM momentum) will approach the constant  value $2 P^2
/3M$ for large $r$. This is precisely the behavior observed in I, for the
particular case of a conformally flat initial data. In the next Section we
explore further the properties of this type of perturbations.

\section{A gravitational memory effect}

Consider again the Zerilli equation (\ref{eqZer1}).  The properties of the
solution obtained by evolution of initial data of compact support have been
extensively studied following the original work of Price \cite{Price}
and Wald \cite{Wald}. In the case of interest in the present analysis,
however, the initial data is only assumed to be smooth and uniformly bounded,
with $\partial \psi /\partial t$ vanishing for $r_* \rightarrow \pm \infty$,
while $\psi$ may approach constant non vanishing values as $r_* \rightarrow \pm
\infty$, and, therefore, the results obtained in the case of compact support are
not immediately applicable. We may resort, nevertheless, to plausible, albeit
non rigorous, arguments to predict the evolution of this type of data. As we
shall see, the results are in agreement with what we obtain by numerical
methods.

First we notice that the ``energy'' integral
\begin{equation}
\label{int01}
{\cal{E}}(t) =\int_{-\infty}^{+\infty} \left[ (\partial
\psi(t,r_*) /\partial t)^2 +(\partial \psi(t,r_*) /\partial r_*)^2
+\psi(t,r_*)^2 V(r_*) \right] dr_*
\end{equation}
is finite for the initial data (at $t=0$) that we are considering, provided
only that $\partial_t \psi \rightarrow  0$ sufficiently fast for $r_*
\rightarrow \pm \infty$.

Moreover, since at large $|r_*|$ the Zerilli equation approaches the free wave
equation form, the limits of $\psi$ for $r_* \rightarrow \pm \infty$ are not
changed by evolution through a finite time. Therefore, ${\cal{E}}(t) $ should be
constant in time, because, on account of (\ref{eqZer1}), we have,
\begin{equation}
\label{int02}
{d \over dt} {\cal{E}}(t) = \lim_{r_* \rightarrow +\infty}
\left({\partial \psi \over \partial t}\right)  \left({\partial \psi \over
\partial r_*}\right)-  \lim_{r_* \rightarrow -\infty}  \left({\partial \psi
\over \partial t} \right)  \left({\partial \psi \over \partial r_*}\right)
\end{equation}
and therefore $d{\cal{E}}(t)/dt=0$, for finite time.

Assume now that $M$ is the black hole mass, and  $K_1$ and $K_2$ are large
positive numbers. We may consider now initial data ${\cal{D}}_0$ that coincides
with our data ${\cal{D}}$ inside an interval ${\cal{I}}=(-K_1 M , K_2 M) $ of
$r_*$,  but is of compact support outside of ${\cal{I}}$. We expect
that the Zerilli function $\psi_0$ resulting from the evolution of ${\cal{D}_0}$
in the domain of dependence of ${\cal{I}}$ will display a ``standard''
behavior, namely, at sufficiently large $t$, we expect $\psi_0$ to display a
quasi-normal ringing wave form, plus a ``tail'', and essentially vanish for
fixed $r_*$, after the passage of the quasi-normal ringing signal. But, since
${\cal{D}}_0$ and ${\cal{D}}$ coincide in ${\cal{I}}$, this should also be the
behavior of the function $\psi$ resulting from the evolution of ${\cal{D}}$ in
the domain of dependence of ${\cal{I}}$.

We may also get an idea of the behavior of $\psi$ at large  $r$, (this is
somewhat simpler than for  $r_*$, because it avoids irrelevant logarithms) if we
assume that at $t=0$ we have $\partial_t\psi=0$, and $\psi$ admits an asymptotic
expansion of the form, (as is true, for instance, for the data of
\cite{KhGlPu}),
\begin{equation}
\label{asynt01}
\psi(r,t=0) \simeq C_0 +C_1/r+C_2/(r)^2 + \dots
\end{equation}
In this case, an asymptotic expansion $\psi(r,t)$, satisfying this initial
data and (\ref{eqZer1}) is of the form,
\begin{equation}
\label{asynt02}
\psi(r_*,t) \simeq C_0 +C_1/r+C_2/r^2 -3 C_0 t^2/r^2+
{\cal{O}}(t^2/r^3)
\end{equation}

Then, for large $r$ we may assume that the ${\cal{O}}(t^2/r^3)$ terms are
negligible for $t < r$, and even for $t \sim r$. If this is correct, for
sufficiently large $t$, irrespective of other details, $psi$ should change sign
around $r \simeq \sqrt{3} t$, and smoothly approach the constant value $C_0$ as
$r \rightarrow +\infty$. Moreover, we find that $\partial_r \psi \simeq 2
C_0/r$, and $\partial_t \psi \simeq 2 \sqrt{3}C_0/r$ at point where $\psi$
changes sign, and therefore both approach zero as $t$ increases. Notice that
 the point where $\psi=0$, ``moves faster than light''. This, however, is only a
consequence of the form of the initial data, and does not imply any causality
violation.

\begin{figure}
\includegraphics[width=6in]{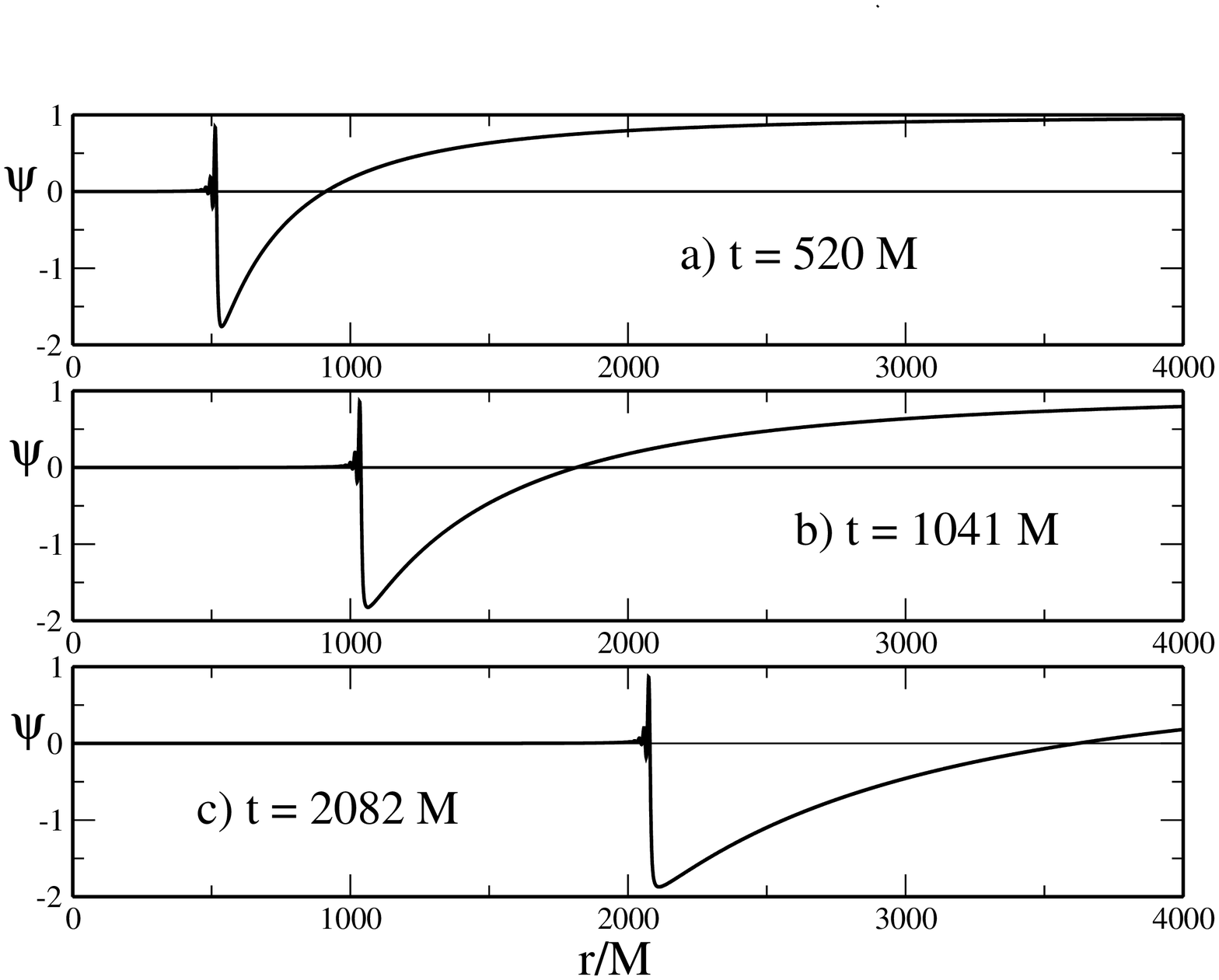}
\caption{\label{fig1}The evolution of initial data of the form $\{\psi(t=0,r)=1,
\psi_t(t=0,r)=0\}$, at times $t=520 M$, (plot a)), $t=1041 M$, (plot b))
and $t=2082 M$, (plot c)). The quasinormal ringing, noticeable as a
structure near $r=500 M$, $r =1000 M$, and $r=2000 M$, respectively in a), b)
and c) is shown enlarged in Figure 2, a), for $t=520 M$. Notice that the point
where $\psi$ first changes sign moves to the right at about twice the speed of
the quasinormal ringing, as indicated in the text}
\end{figure}
\begin{figure}
\includegraphics[width=6in]{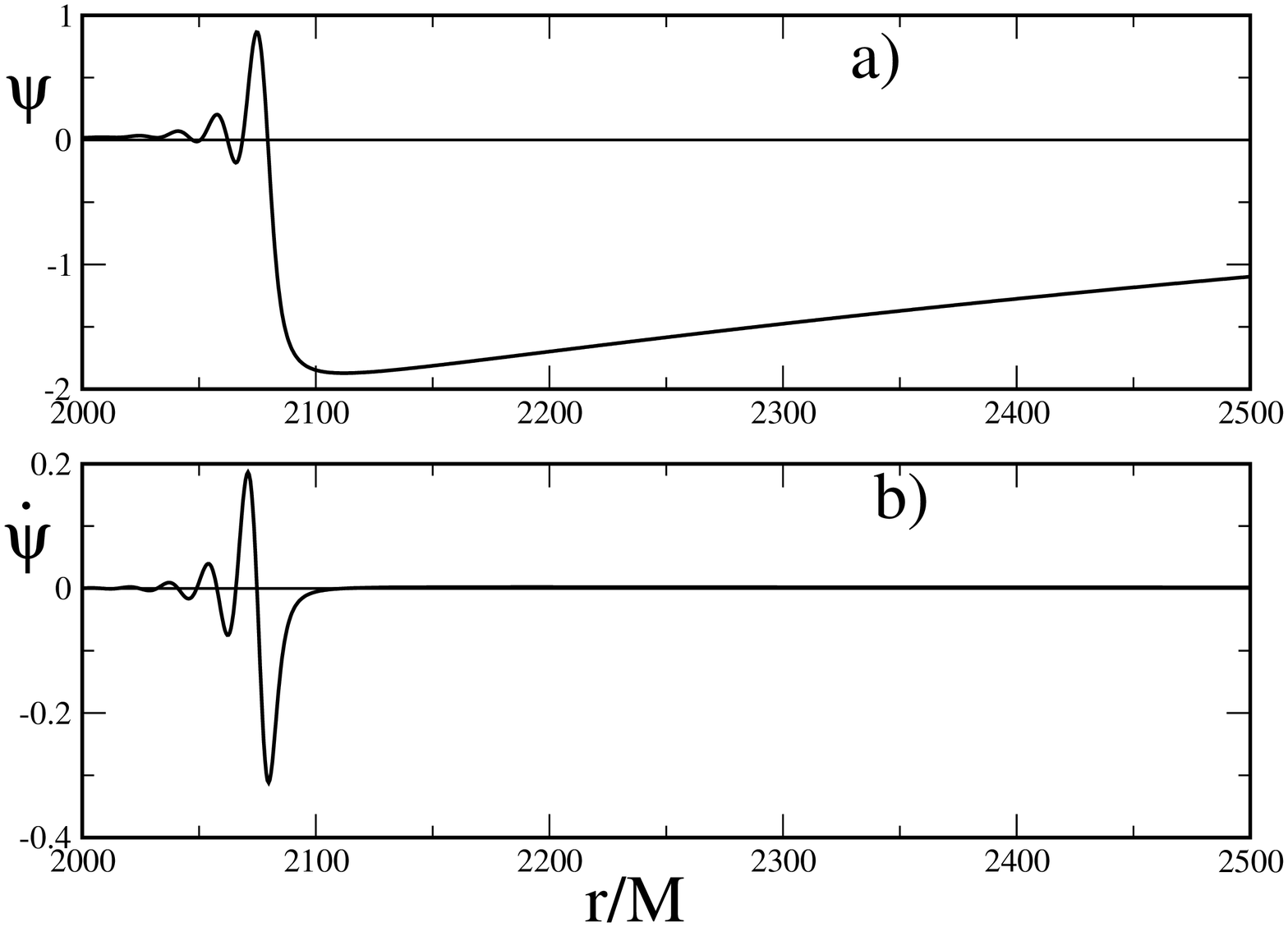}
\caption{\label{fig2} The region where the wave form is dominated by quasinormal ringing, at
$t= 520 M$, is indicated for $\psi$ in a), and for $\partial \psi/ \partial t$
in b) }
\end{figure}

Putting these results together, we expect that after some time $t_0$ of
evolution of our initial data, we should see an essentially vanishing wave form
up to $r \simeq t_0$, where we should see some quasinormal ringing type wave
form, followed by a region where $\psi$ slowly changes sign and finally
approaches the constant asymptotic value of the initial data. This is indeed
what happens if we numerically integrate the Zerilli equation. In Figure
\ref{fig1} we indicate the results of evolving initial data of the form
$\psi(r_*,t=0)=1$, $\partial_t \psi(r_*,t=0)=0$, form $M=1$, through several
times. Curve a) corresponds to $t=520 M$, curve b) to $t=1041M$, and c) to
$t=2082M$. A detail of the quasinormal ringing part, for $t=510M$, is included
in Figure 2, a), while Figure 2, b) displays the time derivative of $\psi$. It
can be seen that these results display all the expected features. In particular,
we find a form of ``gravitational memory effect'', since at large distances from
the black hole, the gravitational wave amplitude $\psi$ starts with a certain
non vanishing constant value and goes to a different constant value (in this
case zero), after some quasi-normal ringing signal is observed. We discuss this
effect in more detail in the next Section.

\section{ Final comments and conclusions}

The results obtained in the previous Section indicate the existence of a certain
gravitational memory effect associated to a particular type of initial data, for
instance, that obtained for single boosted black holes using the Bowen - York
Ansatz \cite{BoYo}. We may picture what this implies by considering its effect
on an interferometric type detector of gravitational waves, placed at a large
distance from the black hole, that is turned on at $t=0$. There one finds a very
slow drift of the equilibrium position, followed by the quasi-normal ringing
signal, with the interferometer ending up in an equilibrium position that is
shifted with respect to the initial one, in an effect that holds certain
similarity to the well known ``gravitational memory'' effect \cite{christo,thorne,blanchet}. In
fact, this is closer to the type of effect originally envisioned by Grischuk,
et.al. \cite{grischuk}, while its relation to the Christodoulou
\cite{christo} type of gravitational memory is unclear to us,
although we remark that the nonlinear nature of the Einstein equations is
involved, since we had to consider second order gauge transformations to arrive
at our final result.

It is well known that the initial data constructed in accordance with the
conformal flatness prescription has the physical drawback that it generally
implies incoming gravitational radiation in the past. For binary black
hole collisions, such as in the Misner \cite{misner} initial data or, more
generally, for boosted black hole data as considered in \cite{boosted},
this seems to introduce no particular undesirable or unexpected features, at
least in the close limit approximation in center of momentum frame. For a
perturbed single boosted black hole, however, we find that the conformal
flatness of the initial data introduces a new type of gravitational memory
effect that we would expect to be absent in the absence of incoming
gravitational radiation. Going back to the original problem of the
``appropriate'' asymptotic behavior of the initial data for a perturbed boosted
single black hole, we may conclude that the ``natural'' choice that would be
physically expected in the absence of substantial incoming radiation, should
lead in the boosted frame where the momentum is $P$, to a Zerilli function that
approaches the constant value $\psi \rightarrow (2/3) P^2/M$ for large $r$ , in
agreement with the discussion carried out in this Paper.

Finally, our results can also be considered in relation with analyses such those
of Kennefick \cite{kennefick}, applied, for instance, to a system such as a
black hole surrounded by a spherically symmetric stationary mass distribution,
that eventually collapses in an asymmetrical manner, inducing a recoil of the
final black hole, together with the emission of gravitational radiation. In
accordance with our analysis, far away from the source, we have initially a
stationary Schwarzschild spacetime, where $\psi$ vanishes, but, after the
passage of the gravitational radiation associated to the collapse, the spacetime
should correspond to a ``boosted'' Schwarzschild, where $\psi$ is a non
vanishing constant, exactly as envisioned in \cite{kennefick}. Notice that the
final value of $\psi$ is related to the recoil momentum as derived in Section
XX, and this in turn is related to the anisotropy of the emmited radiation, so
that we seem to have a further interpretation for the Cristodoulou type of
gravitational memory.

\section*{Acknowledgments}

This work was supported in part by grants of the National University of
C\'ordoba, and Agencia C\'ordoba Ciencia (Argentina). It was also supported in
part by grant NSF-INT-0204937 of the National Science Foundation of the US. RJG
was a visitor at Department of Physics and Astronomy, Louisiana State
University, Baton Rouge, during part of the completion of this work. The
authors wish to thank Jorge Pullin for his helpful comments. The
authors are supported by CONICET (Argentina).


\appendix*

\section{A boosted Schwarzschild metric}

Consider the Schwarzschild metric, written in the form
\begin{equation}
\label{Sch1}
ds^2 = -(1-2M/r) dt^2 + (1-2M/r)^{-1} dr^2 +r^2 [ d \theta^2 +\sin^2 \theta d \phi^2]
\end{equation}
We consider (\ref{Sch1}) only for $r > 2M$, and all $t$. In that
region, we may consider, instead of $t,r,\theta, \phi$, new coordinates
$t',r',\theta',\phi'$, related to the old ones by
\begin{eqnarray}
\label{Sch2Schprime}
\phi & = &  \phi' \nonumber \\
\theta & = &  \theta' -(v t'/r')\sin(\theta') \nonumber \\
r  & = & r' -v t'\cos(\theta') \nonumber \\
t & = &  t'-v r'\cos(\theta')
\end{eqnarray}

This coordinate transformation is motivated by considering a Lorentz boost along
the positive $z$-axis, with velocity $v$, of an auxiliary coordinate system
$t,x,y,z$, related to $t,r,\theta,\phi$ as flat cartesian coordinates to the
corresponding Schwarzschild spherical polar coordinates, and keeping only terms
of order $v$, but this is not central to our discussion. The transformation
(\ref{Sch2Schprime}) can be used to write the Schwarzschild metric in
$t',r',\theta',\phi'$, coordinates. The resulting expression for the line
element is rather long and we shall not display it here. The important point is
that it is diffeomorphic to (\ref{Sch1}), and the metric components are smooth
functions of $v$ near $v=0$, with (\ref{Sch1}) recovered for $v=0$. If we expand
the metric coefficients in powers of $v$, we find that to order $v$ the metric
is asymptotically flat in the sense used in Section IV. We may use (\ref{eqPz})
to compute $P$. The result is $P = vM$, which corresponds to the relativistic
momentum of an object of (rest) mass $M$, computed to order $v$. If we consider
now the expansion to order $v^2$, we find that the metric is {\em not}
asymptotically flat to that order, again in agreement with Section IV. We may
now consider a new coordinate transformation, of order $v^2$, to restores the
asymptotic flatness, but, since the Zerilli function is invariant under this
transformation, we may use the results obtained already in
$t',r',\theta',\phi'$ coordinates to compute it. The result is,
\begin{equation}
\psi_B(t'=0,r') = {4 v^2 M (r'-2M)(3r'+2M) \over 9 r' (2r'+3M)} \simeq - {2
\over 3} M v^2 + {\cal{O}}(1/r')
\end{equation}
which displays the expect asymptotic constant value for $\psi_B \simeq 2 P^2/
(3 M)$.



\begin{thebibliography}{99}

\bibitem{PrPu}  R. H. Price and J. Pullin, Phys. Rev. Lett. {\bf 72} (1994) 3297
.

\bibitem{PuKy} For a recent review, see, e.g., J. Pullin, Prog.Theor.Phys.Suppl.

136 (1999) 107-120

\bibitem{ReWhe} T. Regge, J. Wheeler, Phys. Rev. {\bf 108} (1957) 1063.

\bibitem{Zerilli} F. J. Zerilli, Phys.\ Rev. {\bf D2} (1970) 2141.

\bibitem{physrep} R. J. Gleiser, C. Nicasio, R. H. Price, J. Pullin,
  Phys.Rept. 325 (2000) 41-81

\bibitem{moncrief} V. Moncrief, Ann. Phys. (NY) {\bf 88} (1974) 323.

\bibitem{KhGlPu} R. J. Gleiser, G. Khanna, J. Pullin, Phys.Rev.
{\bf D66} (2002) 024035

\bibitem{Bruni} M. Bruni, S. Matarrese, S. Mollerach, S. Sonego,
Class.Quant.Grav. {\bf 14} (1997) 2585-2606

\bibitem{Price} R. H. Price, Phys. Rev. {\bf D5} (1972) 2419.

\bibitem{Wald} B.S. Kay and R.M. Wald, Class.Quant.Grav.{\bf 4} (1987) 893

\bibitem{BoYo} J. M. Bowen, J. W. York, Phys. Rev. {\bf D21}, 2047 (1980).

\bibitem{christo} D. Christodoulou, Phys. Rev. Lett. {\bf 67} (1991) 1486.

\bibitem{thorne} K.S. Thorne, Phys.Rev. {\bf D45} (1992) 520.

\bibitem{blanchet} L. Blanchet and T. Damour, Phys.Rev. D46 (1992) 4304.

\bibitem{grischuk} V. B. Braginsky and L. P. Grischuk, Zh. Eksp. Teor. Fiz.
{bf 89}(1985) 744 [Sov.Phys. JETP {\bf 62} (1985) 427]

\bibitem{misner} C. Misner, Phys. Rev. {\bf 118} (1960) 1110.

\bibitem{boosted} C. O. Nicasio,  R. J. Gleiser,  R. H. Price, J. Pullin, 
Phys.Rev. {\bf D59} (1999) 044024.

\bibitem{kennefick} D. Kennefick, Phys. Rev. {\bf D50} (1994) 3587.

\end{thebibliography}
\end{document}